# Enhancement in thermoelectric properties of FeSb$_2$ by Sb site deficiency

Anup V. Sanchela[a], Ajay D. Thakur[b,c*], C. V. Tomy [a]

[a] *Department of Physics, Indian Institute of Technology Bombay, Mumbai-400076, India*

[b] *School of Basic Sciences, Indian Institute of Technology Patna, Patna-800013, India*

[c] *Centre for Energy and Environment,*

*Indian Institute of Technology Patna, Patna-800013, India*

(Dated: April 13, 2015)
Original Article/Research

## Abstract

We report a strategy based on introduction of point defects for improving the thermoelectric properties of FeSb$_2$, a promising candidate for low temperature applications. Introduction of Sb deficiency to the tune of 20% leads to enhancement in the values of electrical conductivity ($\sigma$) and Seebeck coefficient ($S$) accompanied with a concomitant suppression in lattice thermal conductivity ($\kappa_{lat}$) values in samples prepared using conventional solid state reaction route. These observations in polycrystalline FeSb$_{2-x}$ provides ample motivation for a dedicated exploration of thermoelectric behavior of the corresponding single crystalline as well as hot-pressed polycrystalline counterparts.

**Keywords**: Thermoelectric materials; Solid state route; deficiency; Thermoelectric properties.

*Corresponding Author
Email: ajay.thakur@iitp.ac.in

Solid state coolers based on thermoelectric materials holds the promise of unleashing the vast potential offered by a wide variety of devices which primarily operate at cryogenic temperatures. These devices, based on manipulating either the spin (spintronic) or the magnetic flux (fluxtronic) offer a multitude of properties far superior than the existing Si- based devices. However, their wide scale usage is hindered due to the complications and costs involved in handling cryogens for reaching the suitable operation temperatures. From this perspective, research in the area of developing thermoelectric materials suitable for low temperature applications is highly desirable [1, 2]. Although the research in the area of thermoelectric materials has primarily been focussed on high temperature applications, there have been several reports on thermoelectric materials suitable for cryogenic temperatures. In particular, Kondo insulators and correlated semiconductors which possess a small gap at the Fermi level (which in turn arises due to the hybridisation of the localized d- or f- states with the broad conduction band) have been explored as promising candidates from this perspective. These include $CeB_6$[3], $CePd_3$[4], $CeRhSb$[5], $FeSb_2$[6–10], $FeSi$[15], $PtSb_2$[16], $YbAl_3$[17], etc, but due to their large thermal conductivity, they are not suitable materials for thermoelectric applications. Among these, $FeSb_2$ has drawn considerable attention recently due to the observation of a very high thermoelectric power of several tens of $\mu V/K$ at 10 K [8, 11–14].

The performance of a thermoelectric material (both as a thermoelectric generator as well as a Peltier cooler) can be described by a dimensionless quantity called the 'figure of merit' ($zT$) such that $zT = S^2\sigma T/(\kappa_e + \kappa_{lat})$. Here, $S$ is the Seebeck coefficient, $\sigma$ is the electrical conductivity, $T$ is the temperature, $\kappa_e$ is the electronic contribution to thermal conductivity and $\kappa_{lat}$ is the lattice (phonon) contribution to thermal conductivity. Although band gap engineering is extensively used in material systems to increase the values of $S$ and $\sigma$, there is a corresponding enhancement in $\kappa_e$ as the value of $\sigma$ is increased [18]. Therefore, suitable reduction in $\kappa_{lat}$ is highly desirable and a number of strategies are employed to achieve it. The value of $\kappa_{lat}$ primarily depends on three scattering mechanisms: (i) phonon-grain boundary scattering, (ii) the phonon-point defect scattering, and (iii) phonon-phonon Umklapp scattering. Whereas, the high temperature behavior is dominated by the Umklapp scattering and the low temperature behavior by the phonon-grain boundary scattering, the phonon- point defect scattering is effective over the entire temperature range. Although conventional micro/nanostructuring methods for enhancing the phonon-grain boundary scattering leads to significant suppression of $\kappa_{lat}$, there is limited improvement in $zT$ due to the concomitant reduction in $\sigma$. Strategies for reducing $\kappa_{lat}$ by introducing atomic scale defects have been employed by numerous groups. In particular, substantial suppression in $\kappa_{lat}$ was observed via introducing lattice imperfections in $Sr_{0.9-x}Ca_{0.1}Si_2$ [19, 41]. Similarly, atomic scale defects introduced in $In_4Se_{3-x}$ lead to substantial suppression in $\kappa_{lat}$ [20]. Nano-structuring based on solution chemistry approach employed by Kieslich et al did not lead to any improvement in thermoelectric properties [21]. Following a nanostructuring approach based on hot-pressing a ball-milled ingot of $FeSb_2$, Zhao et al [22] were successful in reducing the thermal conductivity and thereby improving the $zT$ by a factor of 1.6 at 50K compared to the single crystalline sample. There have been several reports on attempts to improve the $zT$ of $FeSb_2$ via substitution of other chalcogenides including Se, S and Sn at the Sb site and various transition metals at the Fe site [23–25]. It is interesting to see the role played by Sb deficiency on the thermoelectric properties of this very interesting correlated semiconductor $FeSb_2$.

Polycrystalline samples of $FeSb_{2-x}$ (x = 0.0, 0.1, 0.2, 0.3) were prepared by conventional solid state reaction techniques using Fe (powder 99.99%) , Sb (lump 99.99%) were ground in stoichiometric ratio and pressed in to a pellet and placed in to quartz tubes. These quartz tubes were sealed under a vacuum of $10^{-5}$ torr and heated up to 710 °C at the rate of 45 °C/h kept for 24 h and then cooled to room temperature at the rate of 45 °C/h. The sample were reground, resealed, and reheated through identical heating cycle. Powder X- ray diffraction patterns for all samples were collected from 20° to 70° using an X'Pert PRO X-ray diffraction system using Cu-$K_\alpha$ radiation. The Raman scattering measurements were carried using a Jobin Yvon T6400 Raman system in micro-Raman configuration. Argon ion laser of wavelength 514.5 nm as an excitation source. The Seebeck coefficient ($S$) and thermal conductivity ($\kappa$) were measured using the Thermal Transport Option (TTO) of the Physical Property Measurement System (PPMS), Quantum Design (USA) utilizing the two probe configuration. Four probe resistivity ($\rho$) measurements were performed using the resistivity option of the PPMS and Hall coefficient measurements were performed using the corresponding AC Transport Option.

The powder X-ray diffraction (XRD) patterns of the powdered polycrystalline samples of $FeSb_{2-x}$ (x =0 and 0.1) is shown in Fig. 1 (a). The corresponding powder XRD pattern for x = 0.2 sample is shown in Fig. 1 (b). The peaks corresponding to Sb impurity phase corresponding are marked by asterisk. Rietveld refinement was performed using the orthorhombic phase (oP6-$FeS_2$ marcasite type; space group *Pnnm*) for $FeSb_2$ and trigonal phase (space group R3m) for Sb, respectively as the reference structures (using atomic positions from ICPDS databases). Indexing of peaks have accordingly been done. The refinement results for the x = 0.2 sample is shown in Fig. 1(b). The parent $FeSb_2$ sample contains an impurity phase (about 2.7 weight %) comprising of unreacted Sb. A summary of the results of the Rietveld refinement for all the powdered samples of $FeSb_{2-x}$ used in this study are summarized in table 1. $FeSb_2$ has an orthorhombic crystal structure and the variation of lattice parameters with Sb deficiency for our polycrystalline samples FeSb$_{2-x}$ (x = 0, 0.1, 0.2) samples can be found in Table 1 of our manuscript. The lattice parameters *a* and *b* decreases as the deficiency increases up to x = 0.1 and after that increases with the increase in Sb deficiency. Correspondingly, the lattice parameter c increases throughout for deficiency up to x = 0.2. S. J Gharetape et al. [26] reported a similar modulation of lattice parameters with Sb deficiency for FeSb$_2$. Typical sample dimensions are shown in Fig. 2. Densities of the samples compared to the theoretically estimated densities are also marked.

An energy dispersive X-ray analysis (EDXA) of the samples were performed and the results suggested a average nominal stoichiometric proportion of 2.05, 1.92 and 1.84 for FeSb$_2$, FeSb$_{1.9}$, FeSb$_{1.8}$, respectively. However, it should be noted that, EDXA provides a semi-quantitative analysis with detection limits of ~ 0.5 weight % for most elements and therefore the above numbers are indicative and could be taken as a mere indication of successive deficiency introduced at the Sb site. In order to ensure the phase formation, micro Raman spectroscopy was performed and the results for the FeSb$_2$ and FeSb$_{1.8}$ are summarized is Fig. 3. We show a comparative plot of intensity (*I*) versus wave number (Rcm$^{-1}$) for FeSb$_2$ and FeSb$_{1.8}$, in panels (a) and (b) respectively. A multi-peak fitting analysis is performed on the Raman data and a corresponding plot comparing the relative intensities of various modes including $A_g^1$, $B_{1g}^1$, $B_{1g}^2$, $B_{2u}^3$, $B_{3u}^2$ corresponding to FeSb$_{2-x}$ are marked [27]. The modes corresponding to Sb impurity phase are marked by asterisks. Here, it must be emphasized that although Rietveld refinement results suggested an absence of unreacted Sb phase in FeSb$_{1.8}$, nanoscale inclusion of Sb is indicated based on the observed micro-Raman spectra [28]. A detailed transmission electronic microscopy is required to obtain the exact details of unreacted Sb inclusions. The following defect equation is relevant:

$$Fe^{+4}Sb_2^{-4} = Fe^{+4} + Sb_{2-x}^{-(4+x)} + xV_{Sb}^- + xh^+ \quad (1)$$

Here, $V_{Sb}^-$ is the Sb vacancy and $h^+$ represents the hole.

In Fig. 4 (a), we plot $\sigma$ as a function of $T$ for the polycrystalline samples of FeSb$_{2-x}$ (x =0, 0.1 and 0.2) measured using DC four probe technique. The enhancement in the values of $\sigma$ with increase in Sb deficiency is clearly evident. At 100 K, the value of $\sigma$ changes from about 630 S-cm$^{-1}$ for the parent FeSb$_2$ to about 2010 S-cm$^{-1}$ for FeSb$_{1.8}$. Variation of specific heat capacity ($C$) as a function of $T$ for polycrystalline FeSb$_{2-x}$ (x =0, 0.1 and 0.2) is shown in the inset panel in Fig. 4(a). In Fig. 4 (b), we show the variation of $S$ with $T$ for the parent as well as the Sb deficient samples. A five-fold enhancement in the values of $S$ is observed for the FeSb$_{1.9}$ sample compared to the parent material at a temperature of 20 K. Whereas the value of $S$ was ≈-84$\mu$V/K at 20 K for FeSb$_2$, it changes to about ≈-420$\mu$V/K at 20 K for FeSb$_{1.9}$. The inset shows variation of $S$ and $n$ with Sb deficiency at $T$ = 20 K. The carrier concentration decreases from $1.84\times10^{22}$ m$^{-3}$ (for FeSb$_2$) to $1.17\times10^{22}$ m$^{-3}$ (for FeSb$_{1.9}$). As temperature is increased, a change in sign of $S$ is also observed in the temperature range 130 K to 160 K for various compositions of samples with the values of $S$ remaining positive beyond these temperatures. This hints towards the possibility of the presence of both the p-type as well as n-type charge carriers in FeSb$_2$ [29]. All these samples have very similar values of $S$ at high temperatures with the value being ≈23$\mu$V/K at 300 K and ≈34$\mu$V/K at 390K. In Fig. 5, we plot the estimated power factors $S^2\sigma$ as a function of $T$ for the polycrystalline samples of FeSb$_{2-x}$ (x =0, 0.1 and 0.2).

In Fig. 6 (a), we plot the variation of $\kappa$ with $T$ for the samples of FeSb$_{2-x}$ (x =0, 0.1 and 0.2). A $\kappa$ ($T$) maximum of about 2 W/K-m is observed at ~80 K for FeSb$_2$ which is about two orders of magnitude smaller than the corresponding single crystalline samples [30]. A reduced thermal scattering at low temperatures is responsible for the observance of such a maxima. The overall reduction in the magnitude of $\kappa$ ($T$) in polycrystalline samples compared to the single crystalline samples [11, 14] could be attributed to the phenomenon of boundary scattering of phonons and the enhanced Kapitza resistance at grain boundaries [31–33]. In FeSb$_{1.9}$, a $\kappa$ ($T$) maximum of about 2.4 W/K-m is observed at ~50 K. With further increase in Sb deficiency, a $\kappa$ ($T$) maximum of about 0.5 W/K-m is observed at ~65 K. Such a non-monotonicity in behavior as a function of Sb deficiency at low temperatures has been reported earlier [34]. This points to the competing roles of electronic and lattice contribution to thermal conductivity with the increase in Sb deficiency in these samples. A characteristic T$^{3/2}$ behavior of $\kappa$ ($T$) at low temperatures is observed in all the reported samples. It should however be noted that the behavior is clearly monotonic as a function of increasing Sb deficiency at temperatures above 80K. A four-fold decrease in $\kappa$ in FeSb$_{1.8}$ compared to the parent FeSb$_2$ sample at high temperatures demonstrates the role of lattice imperfections in FeSb$_2$ and could be employed as a route for enhancing $zT$ in the system. A knowledge about the variation of electronic mean free path ($l_e$) and phonon mean free path ($l_p$) as a function of $T$ is desirable to understand the underlying physics. Using the kinetic theory result, $\kappa_{lat} = \frac{1}{3} C v_s l_p$, where, $v_s$ is the sound velocity [11]. The lattice contribution to thermal conductivity can be estimated from the thermal conductivity data making use of the Wiedemann-Franz law according to which, $\frac{\kappa_e}{\sigma} = L_0 T$, where the Lorenz number $L_0$ = 2.44 × 10$^{-22}$ W$\Omega$K$^{-2}$ [35]. Thus, $\kappa_{lat} = \kappa - L_0 \sigma T$. Determination of $l_e$ can be done within the Drude approximation where, $l_e = m^* v_e / n e^2 \rho$, with $v_e = \sqrt{3 k_B T / m^*}$, where, $m^* = m_e$, the electronic mass [35]. The temperature variation of $l_e$ and $l_p$ are shown in Fig. 6(b).

A remarkable property of the FeSb$_{2-x}$ (x = 0, 0.1, 0.2) samples is the variation of $\kappa_{lat}$ with $T$ as shown in Fig. 7. The values of $\kappa_{lat}$ have been obtained from the experimentally obtained value of $\kappa$ and $\sigma$. The inset panel show the variation of $\kappa_e$ with $T$ for the samples of FeSb$_{2-x}$ (x =0, 0.1 and 0.2). There is a maxima observed in the values of $\kappa_{lat}$ for the various compositions in the temperature range of ~ 40 K and 80 K. The decrease in the value of $\kappa_{lat}$ below these temperature values indicates a reduction in thermal scattering at low temperatures in these materials. Furthermore, the maximum value of $\kappa_{lat}$ is reduced from about 2 W/K-m for the parent material to about 0.5 W/K-m for $FeSb_{1.8}$. This indicates that the thermal conductivity can be suitably suppressed by the introduction of lattice imperfections. However, here it must also be noted that the values of $\kappa_{lat}$ for $FeSb_{1.7}$ is considerably higher than the parent $FeSb_2$ for a wide temperature range (data not shown here). This in turn indicates that the process of introducing lattice imperfections for reducing $\kappa_{lat}$ cannot be carried out indefinitely. A word of caution is worth mentioning. The measured values of $\kappa$ are very low. According to Quantum Design PPMS specifications, the thermal conductance error is ±5%. In past, Qin et al [36] and Pederson et al [37] reported careful measurements on lower thermal conductivity using the Quantum Design PPMS. We have taken special care while measuring thermal conductivity data.

Arrhenius plot of $\rho$ with $1/T$ for the polycrystalline samples of $FeSb_{2-x}$ (x =0, 0.1 and 0.2) is shown in Fig. 8. Linear fits to data at both the low temperature end as well as the high temperature end are performed for all the samples (see dotted lines in Fig. 8). Estimation of band gaps ($E_{g<}$ and $E_{g>}$ at low and high temperatures, respectively) is obtained from the slopes of these fits. The inset panel in Fig. 8 show the variation of $E_{g<}$ and $E_{g>}$ as a function of Sb deficiency.

Hall measurements on the $FeSb_{2-x}$ (x =0, 0.1 and 0.2) samples were performed to obtain the Hall coefficient $R_H$ and estimates for carrier density ($n$) and mobility ($\mu$) were obtained using $n = 1/e \mid R_H \mid$ and $\mu = \mid R_H \mid /\rho$, respectively. In panel (a) of Fig. 9, we show the variation of $n$ with $T$ for the polycrystalline samples of $FeSb_{2-x}$ (x =0, 0.1 and 0.2). The panel (b) of Fig. 9 shows the corresponding variation of $\mu_H$ with $T$. It can be noted that with the introduction of Sb deficiency (up to x = 0.2), there is an increase in $\mu$ accompanied by a nominal decrease in $n$. FeSb$_2$ is known for its huge Seebeck coefficient in the low temperature range below 100 K [9]. The low temperature resistivity exhibits a thermally activated behavior with a small transport gap. In particular, at low temperatures, it is very difficult to explain decreasing carrier concentration with decreasing resistivity because FeSb$_2$ reveals extreme sensitivity of the carrier concentration to minor change in the sample purity (see [34] for details). At room temperature, the carrier concentration reached a maximum of $5.37 \times 10^{23}$ m$^{-3}$ for FeSb$_{1.8}$ and a minimum of $3.99 \times 10^{23}$ m$^{-3}$ for FeSb$_2$, These results are consistent with resistivity measurements at room temperature. At low temperatures, the resistivity decreases with increasing Sb deficiency. This could be attributed to the increasing charge carrier mobility with increase in Sb deficiency (see [9]). Carrier mobility increases with increasing Sb deficiency in FeSb$_{2-x}$ (x = 0, 0.1 and 0.2) and decreases with increasing temperature. This is due to the increase of carrier concentration with increasing temperatures [9]. As the temperature increases, the thermal vibrations (phonon) increases leading to increased scattering such that the carrier mobility decreases with increasing temperature.

We obtained the scanning electron microscopy (SEM) and tunneling electron microscopy (TEM) images for our samples. Please find below the SEM (see panel (a) and (b) of Fig. 10) and TEM (see panel (c) of Fig. 10) image for the FeSb$_{1.9}$ polycrystalline sample. The TEM image for the polycrystalline sample of FeSb$_2$ is shown in Fig. 10 (d). It can be noted that the TEM images clearly indicate that the grains are crystalline. In addition, the defect boundaries shown between the grains could contribute to the reduction in thermal conductivity due to phonon boundary scattering.

Based on the values of $S$, $\sigma$ and $\kappa$ an estimate of $zT$ is made. In Fig. 11, we show the variation of $zT$ with $T$ for the polycrystalline samples of $FeSb_{2-x}$ (x =0, 0.1 and 0.2) at temperatures below 100 K. The inset shows the variation of $zT$ with $T$ for temperatures up to 390 K. A marked improvement in $zT$ values in Sb deficient samples compared to the parent $FeSb_2$ is observed at 30 K. In addition, a marked improvement in $zT$ is also observed at high temperatures in Sb deficient samples.

In conclusion, an approach based on introducing Sb deficiency for the enhancement of phonon scattering appears to be very effective in reducing the lattice thermal conductivity. This reduction is in addition to the effect due to enhanced grain boundary Kapitza resistance in polycrystalline $FeSb_{2-x}$ samples making it a promising strategy for improving $zT$ for low temperature applications. It is worthwhile to note that we cannot completely ignore the role played by electron-phonon scattering in lowering the thermal conductivity. Electrons participate in lowering the intrinsic thermal conductivity by scattering the surface modes with low energy. The effect of electron-phonon scattering in lowering thermal conductivity is more pronounced at high electron concentrations and low temperatures [38–40]. We believe that reduction of thermal conductivity in the present case is a resultant contribution from a variety of mechanisms including electron-phonon scattering and the phonon scattering by Sb vacant sites, point defects and grain boundaries. It would be interesting to explore the effect of Sb deficiency in single crystalline samples of $FeSb_{2-x}$ and future work in that direction is very relevant. Also, the strategy of introducing Sb deficiency could be coupled with the nano-structuring approach (involving hot pressing/spark plasma sintering of ball milled samples) to attain the ultimate $zT$ for low temperature thermoelectric applications. We hope that our present work will lead to such motivated studies.

CVT would like to acknowledge the Department of Science and Technology for partial support through the project IR/S2/PU-10/2006. ADT acknowledges partial support from the Center for Energy and Environment, Indian Institute of Technology, Patna.

[41] D. Berthebaud, O.I. Lebedev, A. Maignan, Journal of Materiomics 1, 68 (2015).14

TABLE I: Comparison of crystallographic parameters.

| Sample | FeSb$_2$ | FeSb$_{1.9}$ | FeSb$_{1.8}$ |
| --- | --- | --- | --- |
| Nature | Polycrystal | Polycrystal | Polycrystal |
| Data | Powder XRD | Powder XRD | Powder XRD |
| Crystal Structure | Orthorhombic | Orthorhombic | Orthorhombic |
| Space group | *Pnnm* | *Pnnm* | *Pnnm* |
| $a$ ( Å ) | 5.8268 | 5.8255 | 5.8315 |
| $b$ ( Å ) | 6.5338 | 6.5329 | 6.5390 |
| $c$ ( Å ) | 3.1969 | 3.1975 | 3.2007 |
| $V$ ( Å$^3$ ) | 121.710 | 121.693 | 122.054 |
| Rexp % | 5.767 | 3.065 | 3.362 |
| Rpro % | 5.855 | 3.258 | 3.555 |
| Rwp % | 7.511 | 4.151 | 4.488 |
| $\chi^2$ | 1.695 | 1.833 | 1.781 |
| Sb (wt. %) | 2.7 | 0 | 0 |

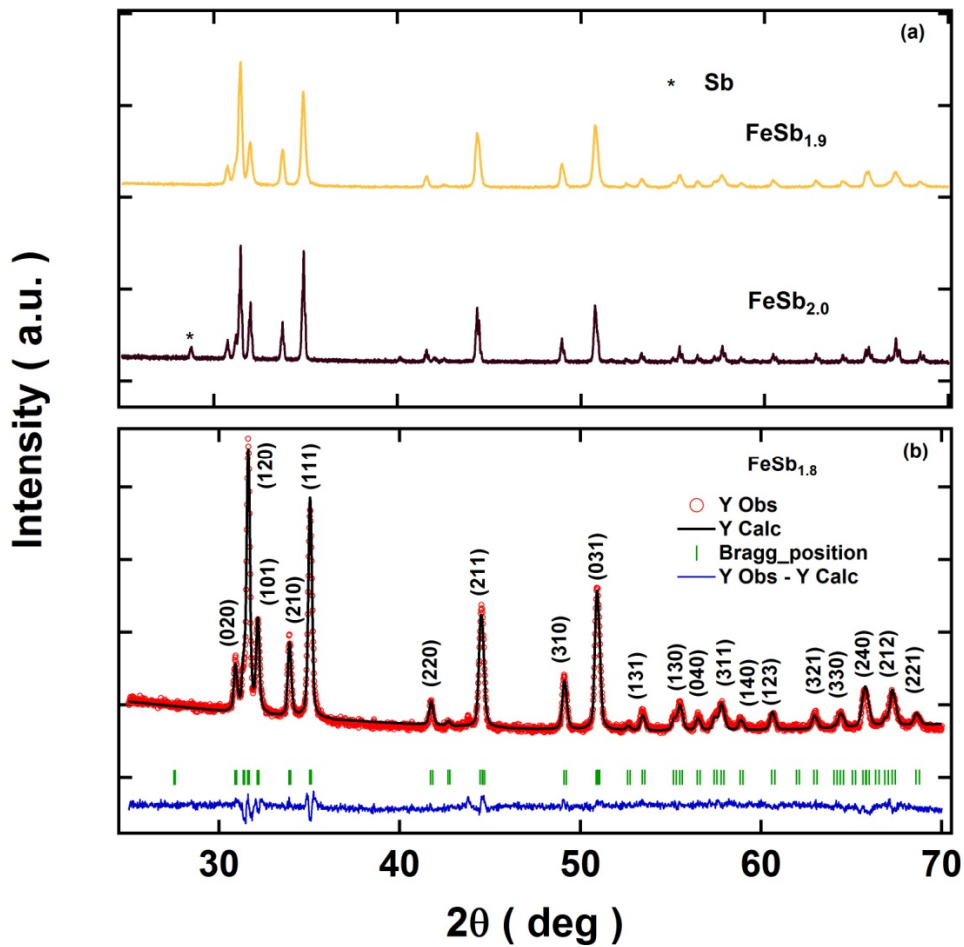

FIG. 1: (Color online) (a) X-ray diffraction pattern for powdered polycrystalline samples of $FeSb_{2-x}$ (x =0 and 0.1). The corresponding powder XRD pattern for $FeSb_{1.8}$ x = 0.2 sample is shown in Fig. 1 (b). Rietveld refinement was performed using the orthorhombic phase (oP6-FeS2 marcasite type; space group *Pnnm*) as reference structures (using atomic positions from ICPDS databases) and indexing of peaks have been done. The refinement results for the x = 0.2 sample is shown in Fig. 1(b). (b) X-ray diffraction pattern for powdered polycrystalline sample of $FeSb_{1.8}$. Also shown is the data for Rietveld refinement performed considering orthorhombic phase (oP6-FeS2 marcasite type structure with space group Pnnm). Peaks have accordingly been indexed.

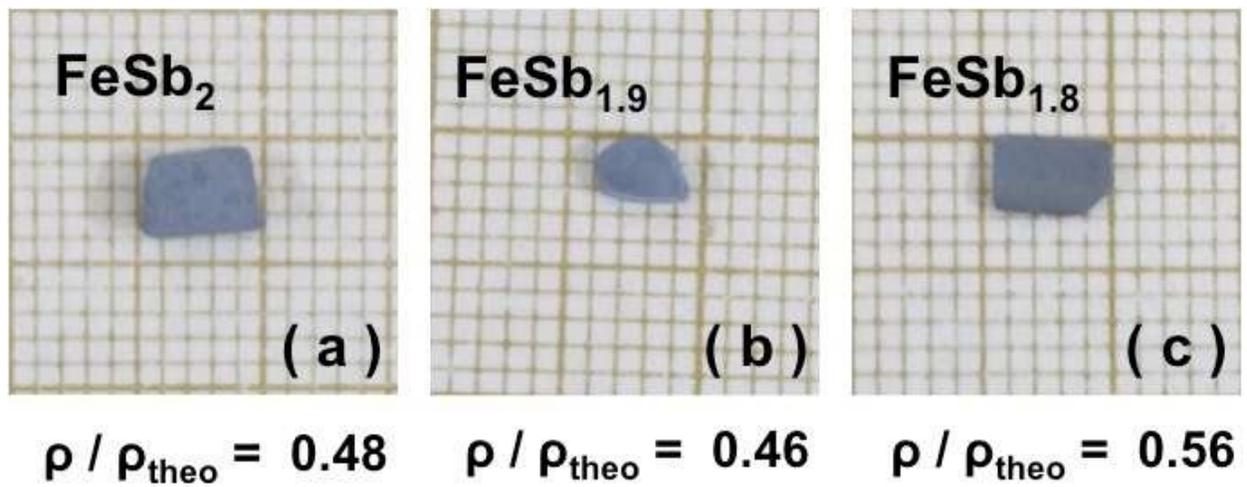

FIG. 2: (Color online) Optical images of the polycrystalline samples used in the present study. The relative densities compared to the theoretically expected densities of the samples are also marked.

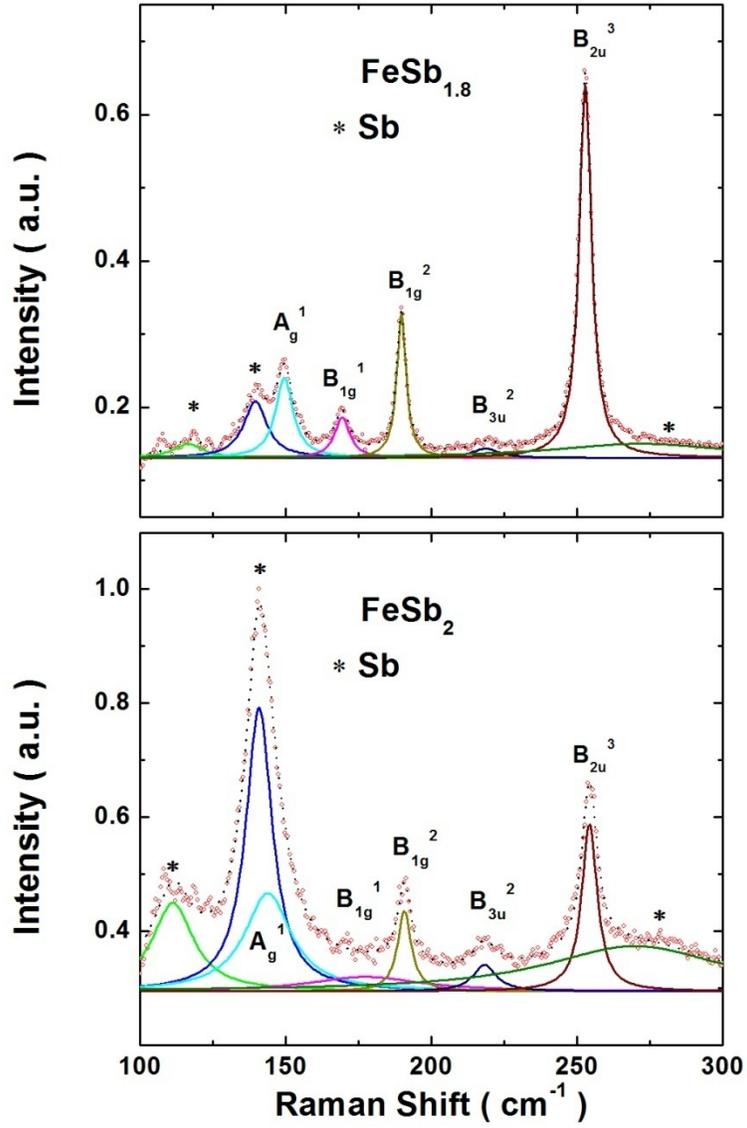

FIG. 3: (Color online) Intensity ($I$) versus wavenumber (Rcm$^{-1}$) for FeSb$_2$ and FeSb$_{1.8}$. The modes $A_g^1$, $B_{1g}^1$, $B_{1g}^2$, $B_{2u}^3$, $B_{3u}^2$ corresponding to FeSb$_{2-x}$ are marked. The modes corresponding to Sb impurity phase are marked by asterisks.

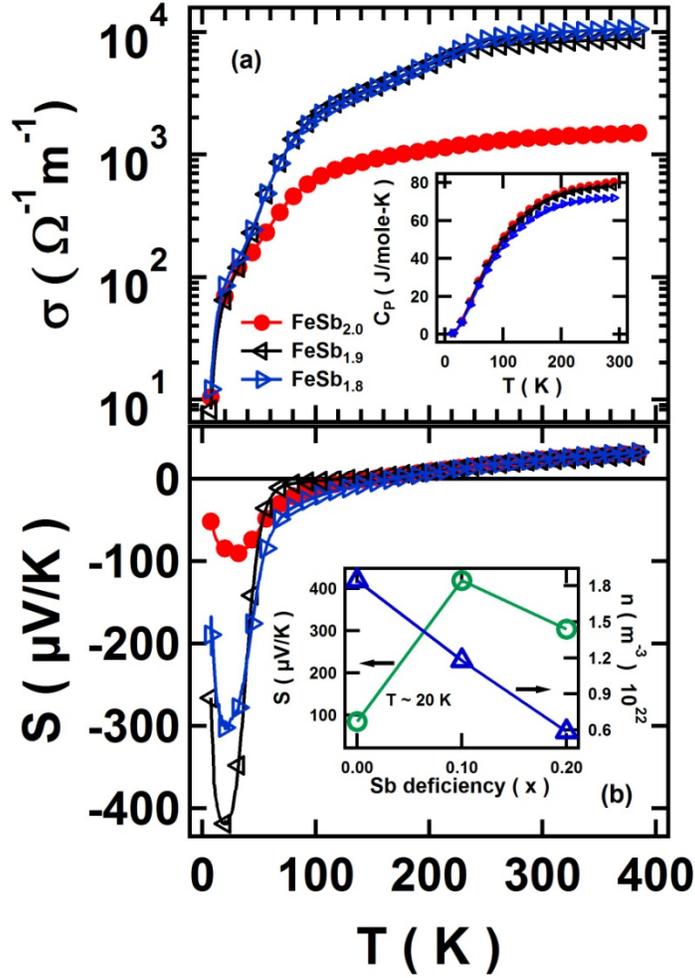

FIG. 4: (Color online) (a) Electrical conductivity ($\sigma$) as a function of temperature $T$ for the polycrystalline samples of $FeSb_{2-x}$ (x =0, 0.1 and 0.2) measured using DC four probe technique. The inset shows variation of specific heat capacity ($C$) as a function of $T$ for polycrystalline $FeSb_{2-x}$ (x =0, 0.1 and 0.2). (b) Seebeck coefficient ($S$) as a function of temperature $T$ for the polycrystalline samples of $FeSb_{2-x}$ (x =0, 0.1 and 0.2). The inset shows variation of $S$ and carrier density ($n$) with Sb deficiency at $T$ = 20 K.

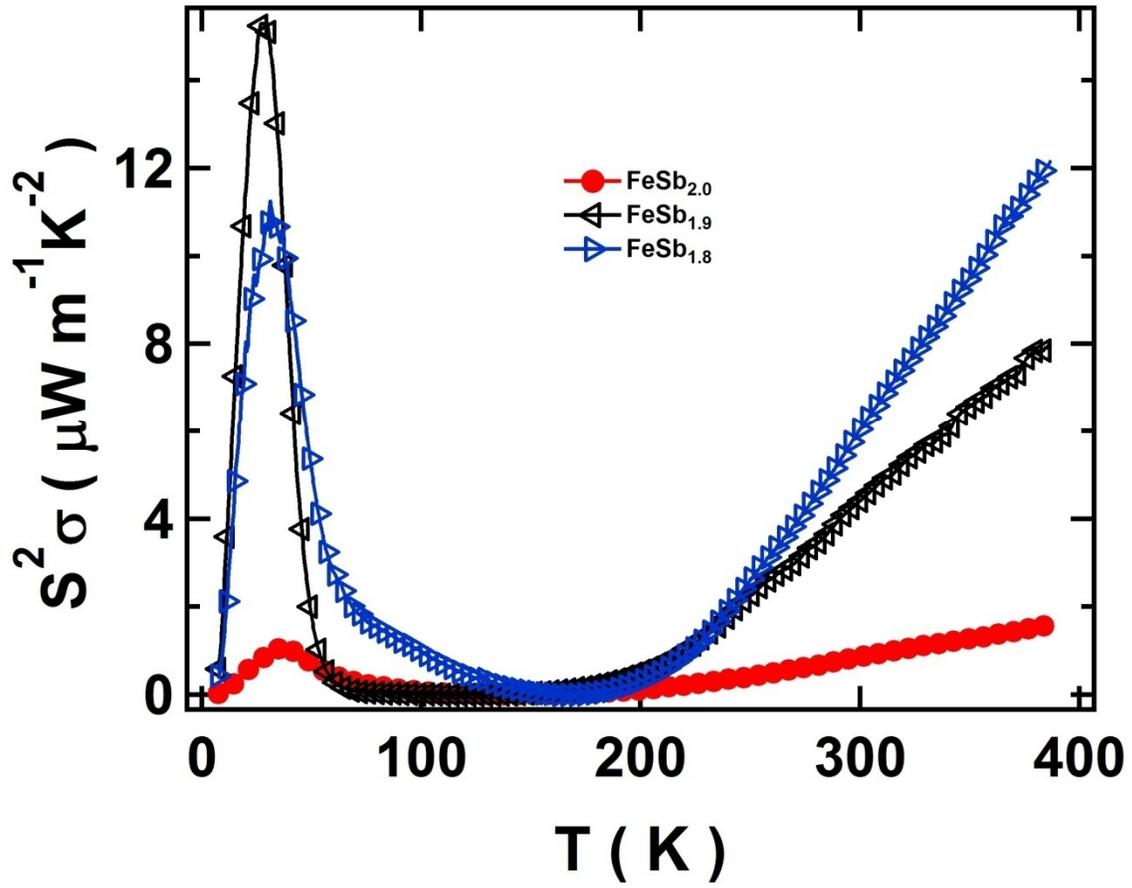

FIG. 5: (Color online) Variation of power factor $S^2\sigma$ as a function of temperature $T$ for the polycrystalline samples of $FeSb_{2-x}$ (x = 0, 0.1 and 0.2).

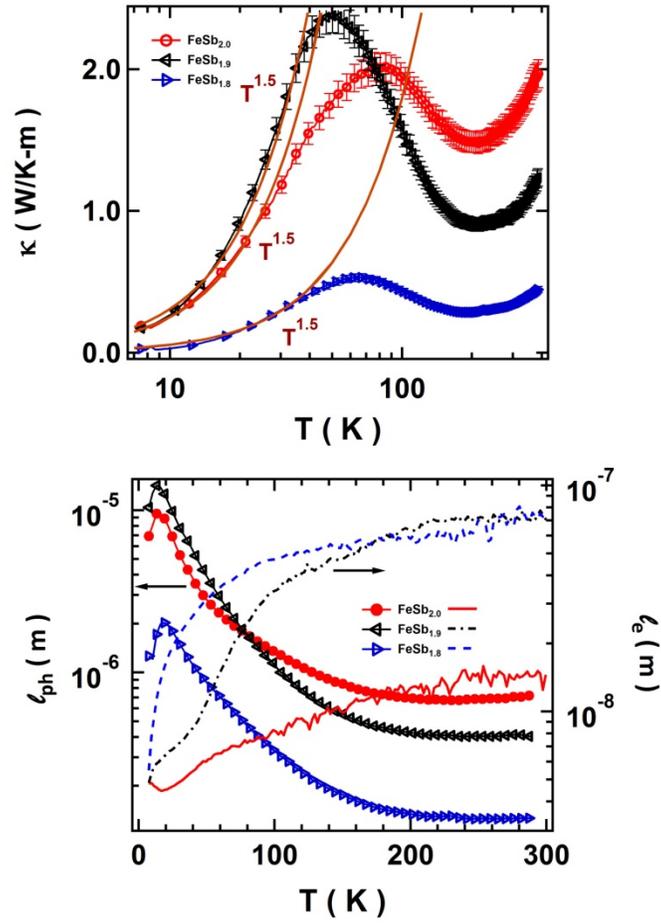

FIG. 6: (Color online) Variation of $\kappa$ and $l_{ph}$ with $T$ for the polycrystalline samples of $FeSb_{2-x}$ (x =0, 0.1 and 0.2).

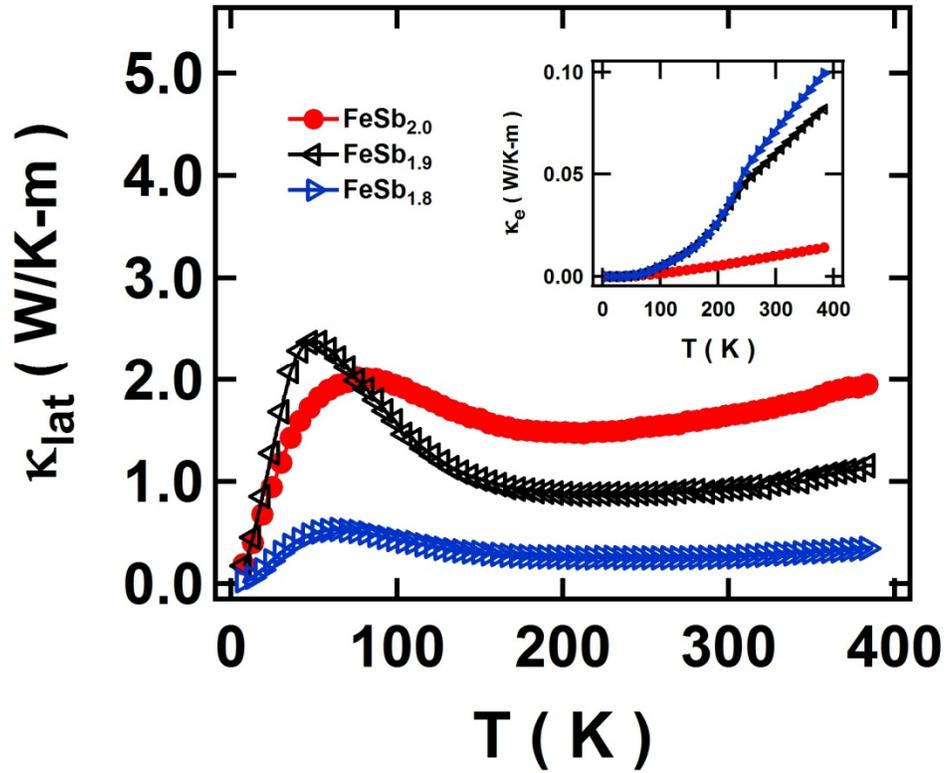

FIG. 7: (Color online) Variation of $\kappa_{lat}$ with $T$ for the polycrystalline samples of FeSb$_{2-x}$ (x =0, 0.1 and 0.2). Inset shows the variation of $\kappa_e$ with $T$ for the samples of FeSb$_{2-x}$ (x =0, 0.1 and 0.2).

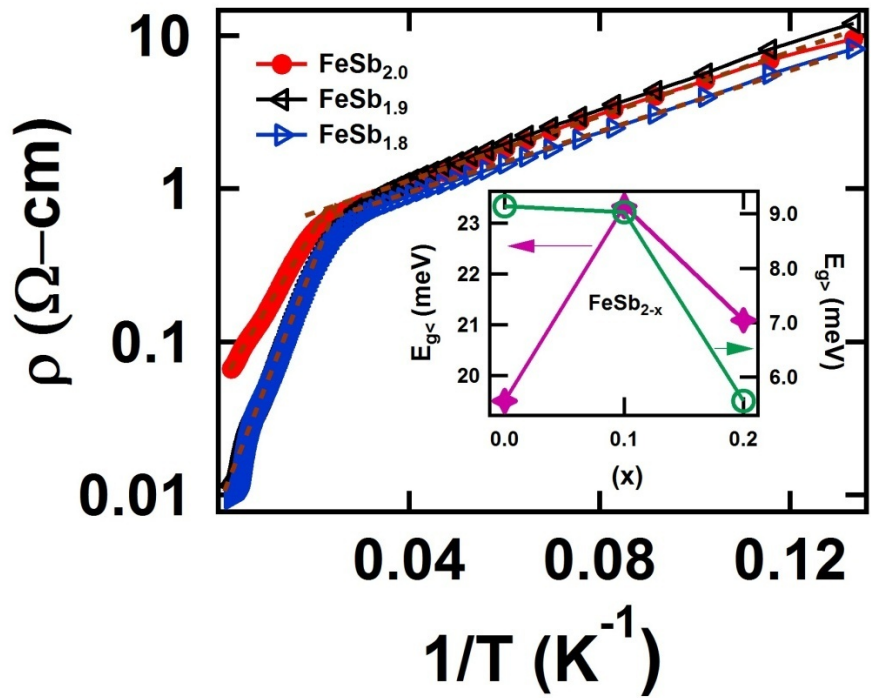

FIG. 8: (Color online) Variation of $\rho$ with $1/T$ for the polycrystalline samples of FeSb$_{2-x}$ (x =0, 0.1 and 0.2). Linear fits to data at the low temperature end and the high temperature end are shown. Inset panel shows the variation of energy gaps with Sb deficiency at low temperatures and high temperatures.

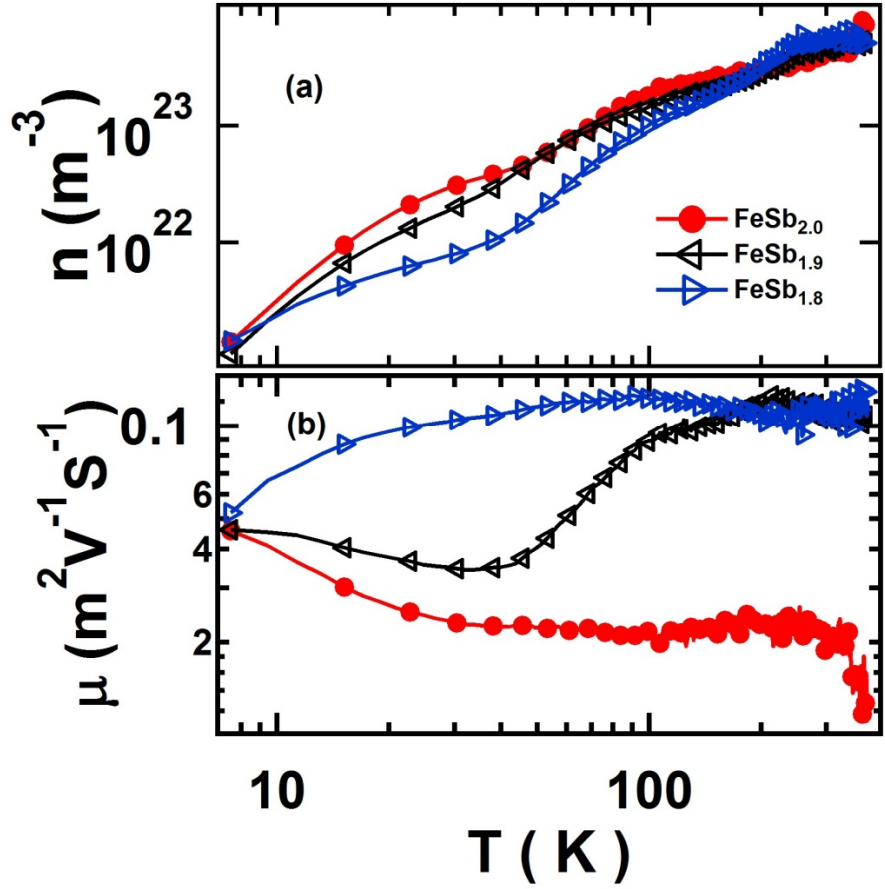

FIG. 9: (Color online) (a) Variation of $n$ with $T$ for the polycrystalline samples of FeSb$_{2-x}$ (x =0, 0.1 and 0.2). (b) Variation of $\mu_H$ with $T$ for the polycrystalline samples of FeSb$_{2-x}$ (x =0, 0.1 and 0.2).

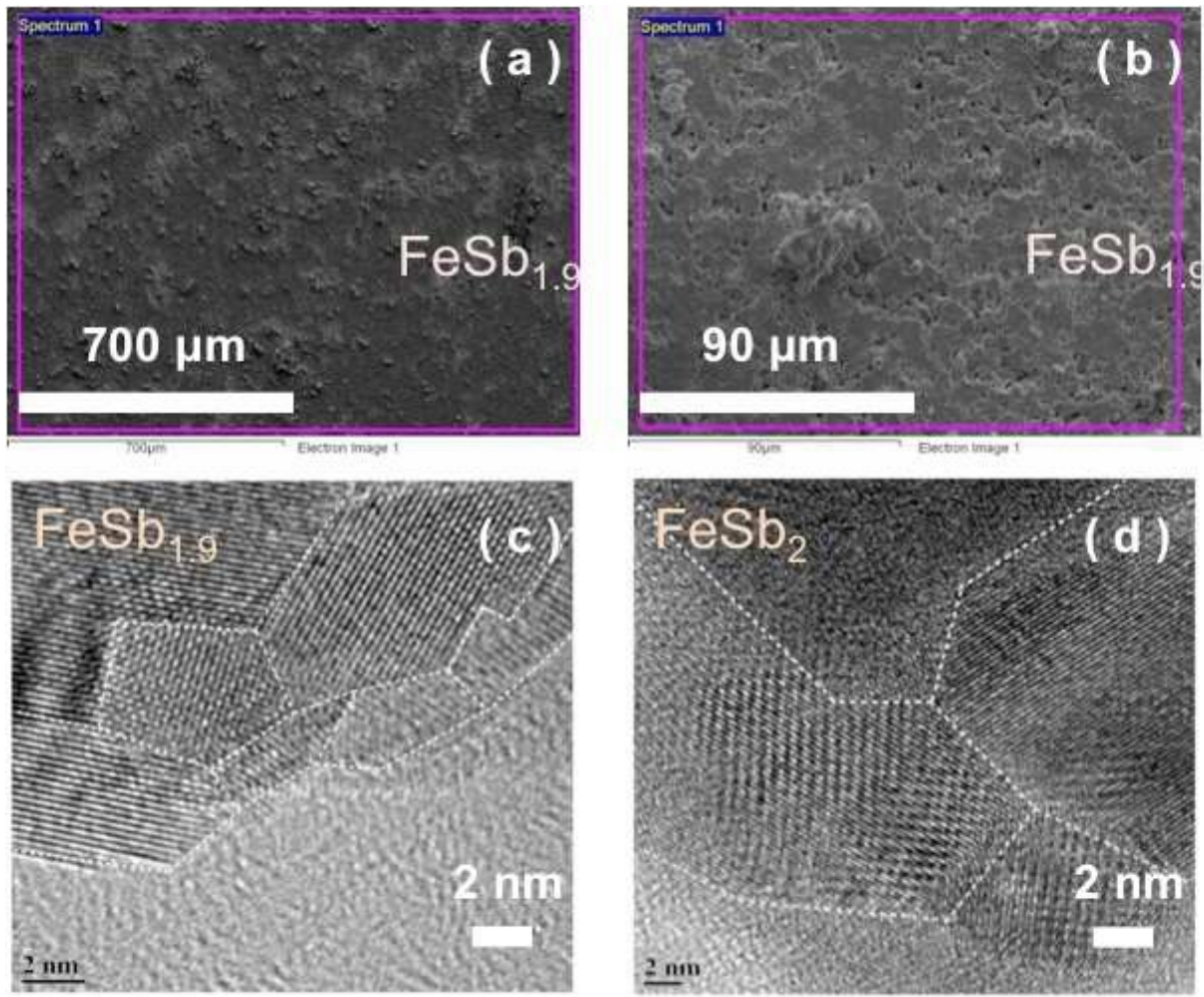

FIG. 10: (Color online) SEM (panels (a) and (b)) and TEM (panel (c) and (d)) images of the polycrystalline samples.

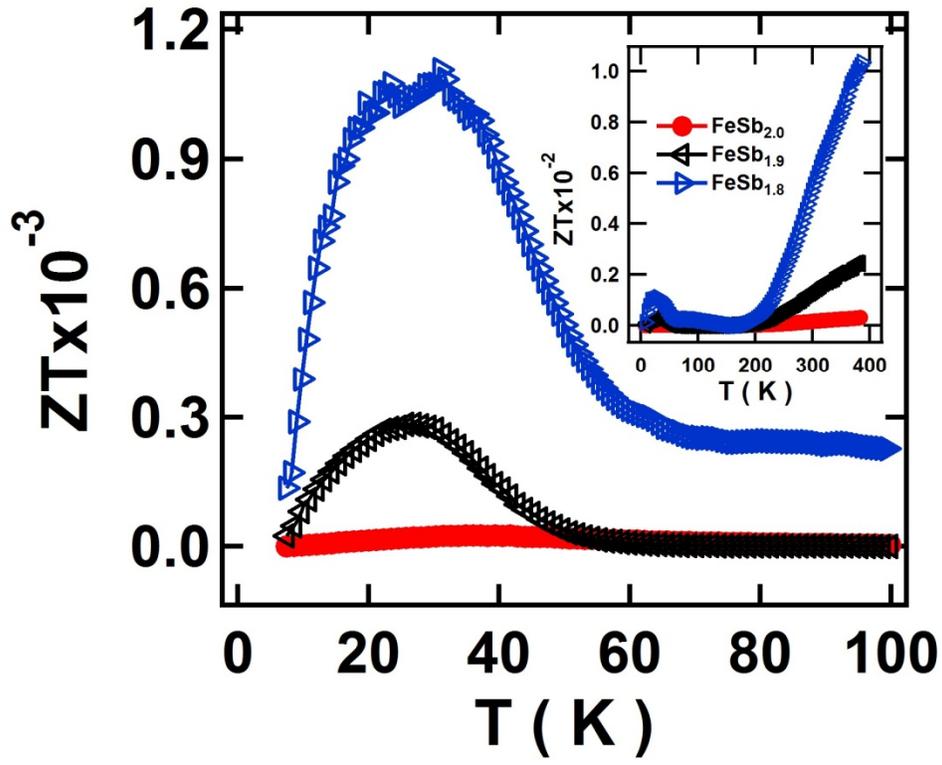

FIG. 11: (Color online) Variation of $zT$ with $T$ for the polycrystalline samples of FeSb$_{2-x}$ (x =0, 0.1 and 0.2) at temperatures below 100 K. The inset shows the variation of $zT$ with $T$ for temperatures up to 390 K.